\documentclass{elsart}
\usepackage{graphics}

\begin{document}

\begin{frontmatter}
\title{Power-law distribution in Japanese racetrack betting}
\author{Takashi Ichinomiya}
\ead{miya@nsc.es.hokudai.ac.jp}
\address{
 Nonlinear Studies and Computation, Research Institute for Electronic
 Science, Hokkaido University, Sapporo 060-0812, Japan.
}
\begin{abstract}
Gambling is one of the basic economic activities that humans indulge in. An
investigation of gambling  activities provides deep insights into the
  economic actions of people and sheds lights on the study of
 econophysics.
 In this paper  we present an analysis of  the distribution of the final
 odds of  the races organized  by the  Japan Racing Association. The
 distribution of the final odds $P_o(x)$ indicates
 a clear power law $P_o(x)\propto 1/x$, where $x$ represents the final
 odds. This  power law can be explained on the basis of the assumption
 that  that every bettor bets his
 money on the horse that appears to be the  strongest in a race. 
\end{abstract}


\end{frontmatter}

\section{Introduction}

 Since Pascal proposed the theory of probability, the phenomenon of
 gambling have given inspirations to many mathematicians,
 physicists and economists. 
 For example, economists have studied  gambling activities in order  to
 examine the market efficiency. If bettors try to maximize their
 expected rewards, it would be  equivalent for  every race and every horse.
 In this case, the gambling market is efficient, and there is no assured
 way to gain money. Gabriel and
  Marsden studied  market efficiency in British racetrack betting and found
  this gambling market to be  inefficient\cite{Gabriel}.  Russo, Gandar
 and Zuber conducted a similar study  on the National Football
 League betting market\cite{Russo}. They found no clear indication of
 breakdown of  efficiency. These results are re-examined Cain {\it et
 al.} and Ioannidis and Peel by different methods\cite{Cain,Ioannidis}

 Econophysics and financial engineering are other examples of 
 studies that are  closely related to gambling.  It is true that
 gambling and trading differs. Though the stock value is  determined
 by the action of traders, the bettors cannot determine the horse who
 win the race. However, the trading of stocks bears some similarity
 to gambling. In trading, traders speculate on
  stocks, attempt to predict which stock value will
 rise in the future. This is similar to gambling, in which  bettors
 attempt to predict. Especially, the rewards of winners is determined by
 the decision of all bettors. This is similar to trading, in which the
 stock value is determined by the action of traders.
 Thus, the investigation of gambling will contribute to the
 advancement of econophysics.  In this regards, Park and Domany reported
 the distribution of odds in Korean horse races\cite{Park}. They found
 the power-law distribution of the final odds. To explain this
 distribution, they proposed the model of betting, in which the bettors'
 estimation of winning probability is affected by the odds. 

 In this paper, we present an investigation of  the distribution of the 
 odds in horse racing in Japan, where national horse races are
 organized  by the Japan Racing Association(JRA).  The form of betting
 in horse race in Japan is
 same as the one in Korea.  There is
only one form of betting: pari-mutuel tote bet.  In this betting system,
 the management expenses are deducted  from the total amount of the bet, and 
 the winners divide the remaining amount among themselves. In this
 paper, the odds is defined as the ratio of the reward  to the bet. If
 we neglect the management expenses, the odds of a horse is given as
 total amount of the bed divided by amount of the bed on the horse. 

 We show that  the distribution of the final odds shows
 power law $P_o(x)\propto x^{-1}$,
 where $x$ represents the final odds. This is different from the one
 obtained by Park and Domany, $P_o(x)\propto x^{-1.7}$.
 We propose the model to account
 for this power-law distribution. One interesting point of this model is
 that this power-law behavior is obtained by assuming bettors to be
 irrational. Here the word `irrational' means that bettors do not try to
 maximize their expected rewards. If we assume bettors to be rational,
 i. e. attempt to
 maximize expected rewards, we obtain power-law $P_o(x)\propto x^{-2}$,
 which is different from empirical data. 

 The outline of this paper is as follows. In the next section, we
 present the distribution of the final odds in horse races organized by
 the JRA. We show  that the distribution of the final odds $P_o(x)$
 and that of winners' odds $P_w(x)$  exhibit power-law behaviors,
 $P_o(x) \propto x^{-1}$ and $P_w(x) \propto x^{-2}$. In section
 \ref{model}, we propose the model to account for this power-law
 distribution. In section \ref{conclusion}, we make some discussion on
 our results.

\section{Investigation of the final odds in racing organized by the
 JRA}

 We analyze the data  of 1750 races organized by  the JRA between
 February  and July 2005. A total of  24493 horses participated in
 these races. For simplicity, we  investigate only the odds of the win bets,
 without  considering  other bets such as dual forecasts or place bets.

In Fig.1, we plot the distribution  of the final odds $P_o(x)$ for each
 month, where $x$ represents the final odds.
 It should be noted  that this distribution is not one of  the
 winning payouts; this figure shows the distribution of the final odds,
 including  the losers' odds.
The distribution $P_o(x)$ for  each month is very similar. At $x<100$,
 $P_o(x)$ follows the power-law distribution, $P_o(x)\propto
 x^{-\gamma}$, 
 and decays exponentially at $x>100$.
 In order  to obtain a clearer picture of  the power-law distribution,
 we plot the
 total half-year distribution  in Fig.2. This data indicates that 
  $\gamma \sim 1$. 
 The distribution of the final winners' odds  is also interesting. In
 the same figure, we also plot the half-year distribution of winners'
 odds.
 This data shows that the distribution of the 
 winning payouts also follows the power law,
 $P_w(x) \propto 1/x^2$.

  It should be noticed that these data
 appear to be   consistent with the efficiency of the gambling market.
 We assume that we bet on a horse whose final odds are represented by
 $x$. Then, the probability that the horse win is proportional
 to $P_w(x)/P_o(x)$. Because the expected reward  is represented by 
 $x P_w(x)/P_o(x)$, the empirical data shows that $P_w(x)/P_o(x)$ is
 proportional to $x$ and that the expected reward is independent of $x$.

\begin{figure}
\resizebox{.45\textwidth}{!}{\includegraphics{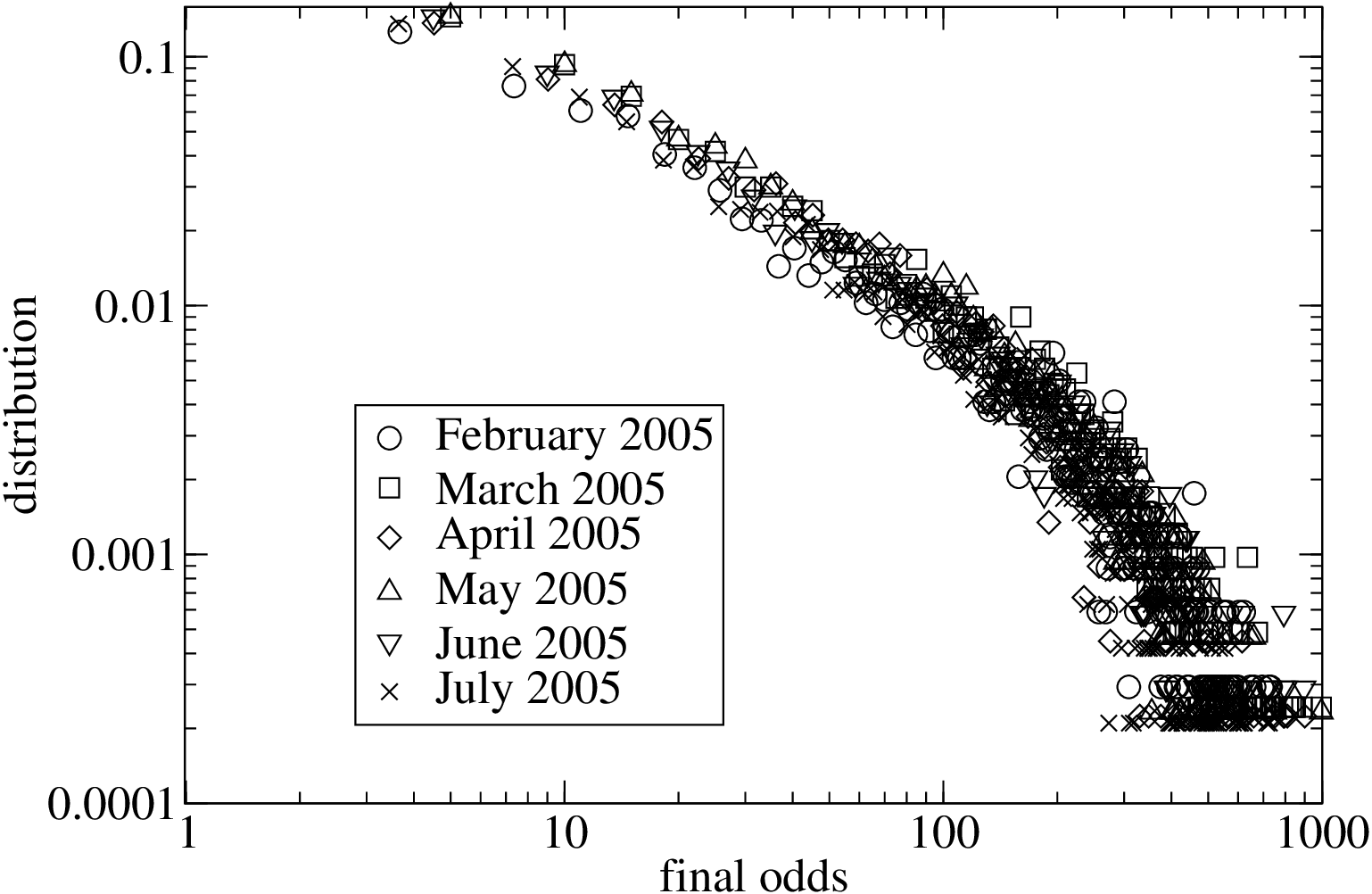}}
\caption{Distribution of the final odds between February and July 2005.\label{fig1}}
\end{figure}
\begin{figure}
\resizebox{.45\textwidth}{!}{\includegraphics{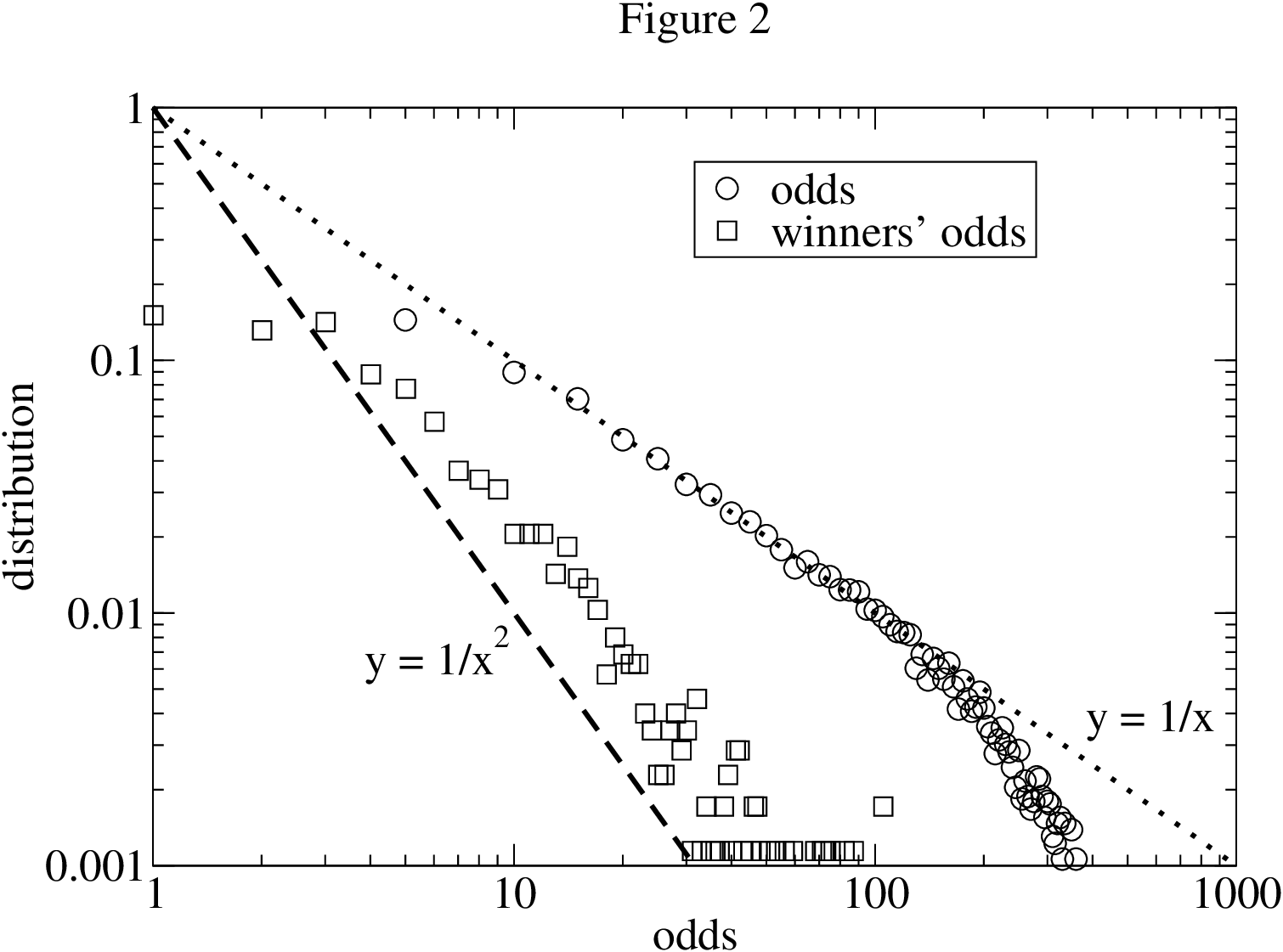}}
\caption{Total distribution of the odds and the winners' odds between
 February and July 2005.\label{fig2}}
\end{figure}

 However,  the efficiency of the gambling market does not mean the
 rationality of bettors. If bettors are rational, they will bet on the
 horse whose expected rewards is maximum, and the
 expected rewards will become equivalent for all horses. It is
 true that rationality of  bettors is a sufficient condition for the efficient
 market, however, it is not a necessary condition.
 It should  also be  noted that the  power-law distribution of
 the final odds cannot be explained solely by the rationality of 
 bettors. We will need some other assumptions to explain the
 distribution. In the next section, we propose a model to account for
 these power-law distributions of the final odds. 

\section{Model simulation\label{model}}

 In this section,  we propose a  model to account for the
 distribution of the final odds. In order to construct the model, we
 must focus on  the following
points. First, the strength of each horse running in  a race varies. One horse
may be stronger than the other horses, and  the weakest  horse may perhaps lose
 the  race. Therefore, we must include the
differences in the  strengths of horses in the model. Second, the strength of
each horse provides the probability of winning. Even the strongest horse can
lose the race if he is in poor health. Therefore, it appears natural to
assume that the probability of winning is given by a function of the strength
of the horse. Third, bettors are unaware of the exact strength of the
 horses.
 While they do know that some horses are  stronger than the others, they
do not know the exact strength and cannot precisely estimate the probability of
winning.

Considering these points, we propose the following model:
\begin{description}
 \item[(1)] Each horse running in a race has a parameter called
	    `strength' $s$, $0 < s \le 1$. $s_i$, the strength of horse $i$, is
	    given by a uniform random number between 0 and 1.
 \item [(2)]Each bettor estimates the strength of horse $i$ as $s_i +
	    r_i$, where $r_i$ is the Gaussian random number. $r_i$ varies for
	    each bettor, and the strength estimated by each bettor also differs.
 \item [(3)]Each bettor bets his money on the horse that appears to have the
	    maximum strength in a race. As observed, bettors do not know the exact
	    strength of the horses. They place their bets under incorrect
	    estimation of the strength given by the previous rule. For
	    simplicity, we assume that every bettor bets the same amount of
	    money.\label{rule3}
 \item [(4)] After each betting, $x_i$, the odds of a horse $i$, is updated as
	    $x_i = n_{tot}/n_i$, where $n_{tot}$ and $n_i$ represents the
	    total number of bettors and number of bettors who bets on
	    horse $i$, respectively. This rule
	    means  that we neglect the management expenses, and the total
	    amount of bets is divided among the winners. This simplification
	    does not lead to any problems.
 \item [(5)]The winning probability of a horse is proportional to his
       strength.
\end{description}

 This model has three parameters. The first parameter is the number of
 horses that participate in one race. Although the number of horses
 differs with each race, in the following simulation, we assume  that 14
 horses participate in one race. This approximates
 the average number of  horses participate in one race, 13.996. 
 The second parameter is the number of bettors. We set this number as
 10000. The qualitative result of this simulation does not depend on the
 details of this parameter, if it is  sufficiently large.   
 The third parameter is $\sigma$, the dispersion of a  random number $r_i$.  In
 our model,  $\sigma$ is the only parameter we should adjust for fitting.

\begin{figure}[t]
 \resizebox{.4\textwidth}{!}{\includegraphics{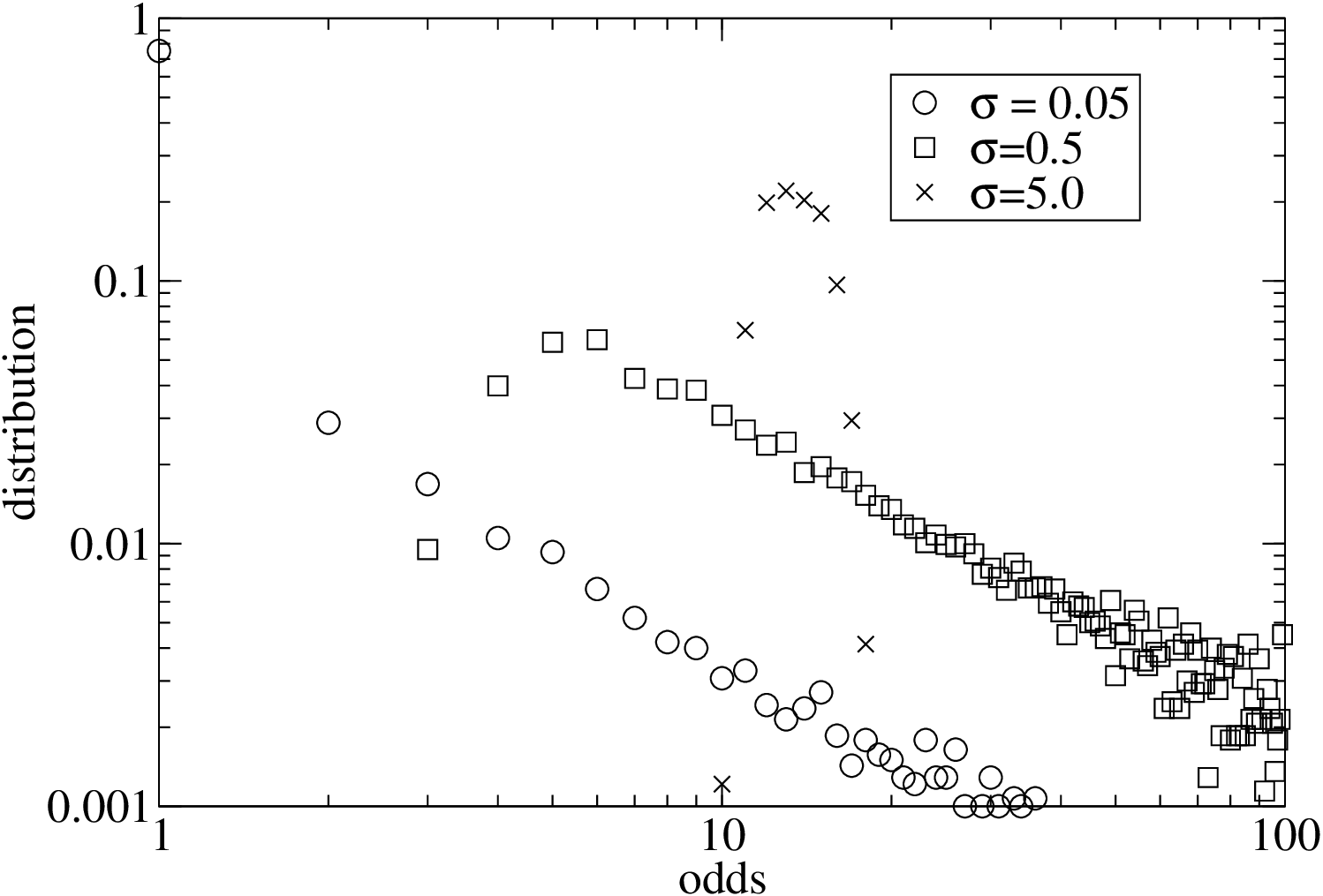}}
 \caption{ Distribution obtained from our model at $\sigma=0.05$, $0.5$
 and $5.0$. Each distribution is obtained by simulating our model 1000 times. 
\label{sigmadep-Po}}
\end{figure}

 To see the $\sigma$-dependence of the distribution of the final odds,
  we plot the distribution $P_o(x)$ obtained by our model for some
  typical $\sigma$ in Fig.\ref{sigmadep-Po}. If $\sigma \ll 1$, almost
  all bettors bet on the strongest horse, and $P_o(x)$ shows sharp peak
  at $x=1$.  On the other hand, if $\sigma \gg 1$ the strength of horse
  scarcely affects the choice of bettors. Every bettor bets on the horse
  almost at random, and $P_o$ will show sharp peak at $x=h_{tot}$, where
  $h_{tot}$ is the total number of horses.
  The results of simulation at $\sigma=0.05$ and $\sigma=5.0$  shown in
  Fig.\ref{sigmadep-Po} are consistent with this qualitative discussion.
  Though the region where  $P(x) \propto 1/x$ appears in both cases,
  these results do not qualitatively coincide  with   empirical data.
 In the following of  this paper, we use $\sigma=0.5$ and compare the
 empirical data and  the result of simulation. 

\begin{figure}
\resizebox{.4\textwidth}{!}{\includegraphics{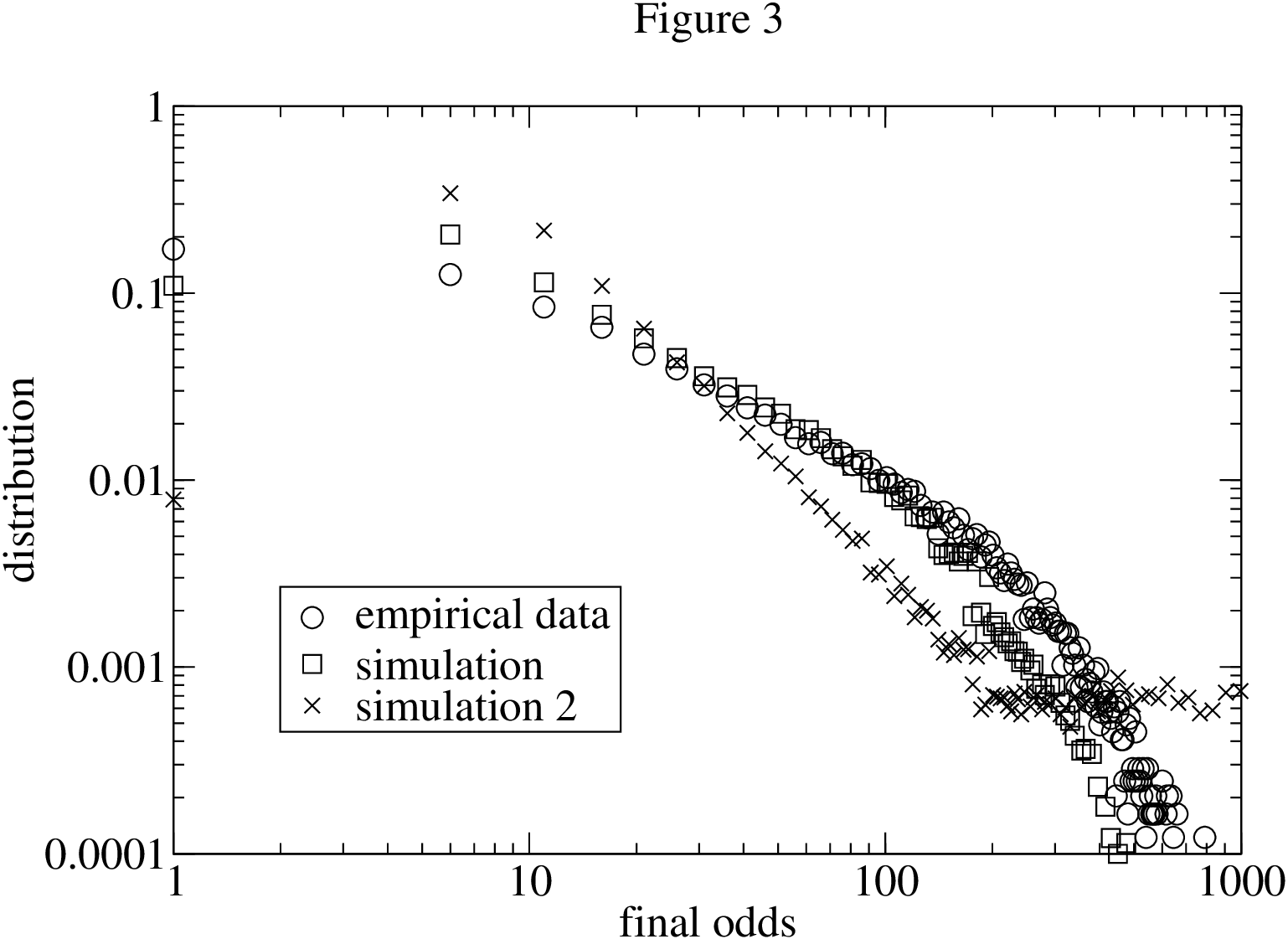}}
\caption{Comparison of the empirical distribution of the final odds and
 model calculation. Circles, squares and crosses
 represent the empirical data, the result of the model simulation and
 the result of the  model simulation in which bettors attempt to
 maximize their expected
 rewards, respectively.\label{fig4}}
\end{figure}

\begin{figure}
 \resizebox{.4\textwidth}{!}{\includegraphics{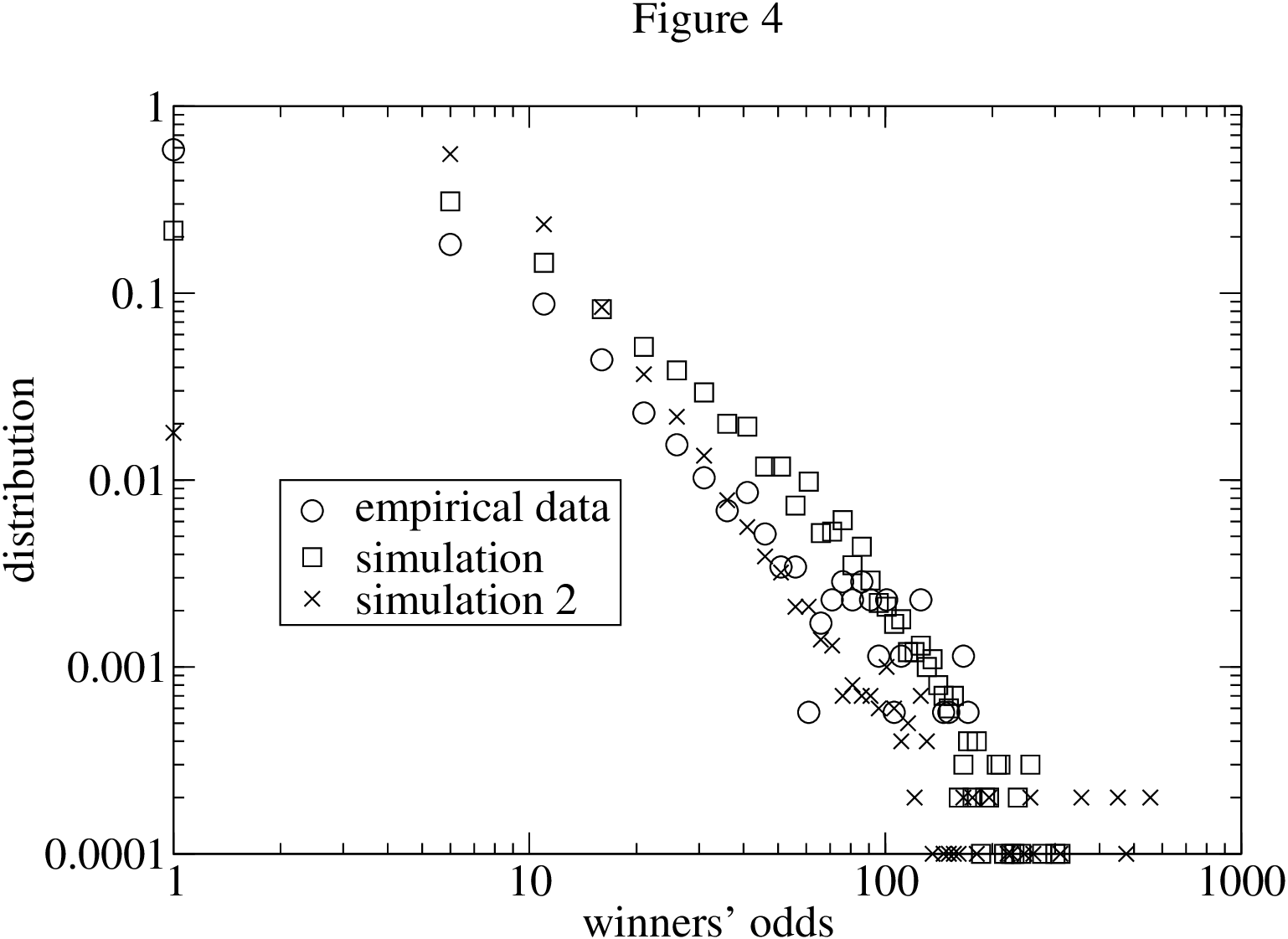}}
\caption{Comparison between the empirical  distribution of the winners'
 odds and model calculation. Circles,  squares and crosses
 represent the empirical data, the result of the model simulation, and
 the result of the  model simulation in which bettors attempt to
 maximize their expected rewards, respectively.\label{fig5}}
\end{figure}

 We calculate the distribution of the final odds at $\sigma=0.5$ by
 simulating this model 10000 times.
 The results of the simulation are shown in Figs. \ref{fig4} and \ref{fig5}.
 With regard to  the distribution of the final odds shown in
 Fig.\ref{fig4}, both the
 power-law behavior at $x<100$
 and the exponential decay at $x>100$ are reproduced well in  our model
 and are  represented by squares.  On the other hand, the power-law
 behavior of the distribution of the
 winners' odds shown in Fig.\ref{fig5} is also
 reproduced by the simulation, although the absolute value differs
 slightly. These results suggest the validity of our model.

 However, the economists who believe in the rationality of people
 will expect that the bettors attempt to maximize their expected
 returns. If this is true, the bettors
 bet their money on the horse whose expected rewards (winning probability
 multiplied by odds)  is maximum, instead
 of betting on the one whose probability of winning is maximum.
 However, the simulation shows that  the empirical data cannot be
 accounted for by using  the `maximize the expected rewards' model.
 In the `maximize the expected rewards' model, we use the following rule
 instead of the rule (3) in our model:
\begin{description}
 \item [(3')]Each bettor bets his money on the horse that appears to have the
       maximum expected rewards in a race. Because the probability that a
       horse wins is proportional to his strength, the expected reward
       of horse $i$ seems proportional to $(n_{tot}+1)(s_i +
	     r_i)/(n_i+1)$. Bettors
       bets his money on the horse $i$ that seems to have maximum  expected
       reward. 
\end{description}
 Here we note that the expected rewards is not $n_{tot}(s_i +r_i)/n_i$,
 because if the bettor bets on the horse $i$, the odds of the horse
 changes from $n_{tot}/n_i$ to $(n_{tot}+1)/(n_i+1)$.
 In this model, the distribution of the odds is not $\delta$-function even
 if $r_i=0$, because bettors bet on the weak horse with high odds. If
 the number of bettors is large enough, the expected rewards $x_i s_i$ will be
 same for all horses. 
 The result of the
 simulation based on this model is indicated by the  cross in
 Figs. \ref{fig4} and \ref{fig5}. In this `maximize the expected rewards'
 calculation, we  take $\sigma=0.001$.
 In this model, the distribution of the final odds $P_o(x)$ and winners'
 odds $P_w(x)$ are approximately proportional to $1/x^2$ and $1/x^3$,
 respectively. These results are inconsistent with the empirical data.

\section{Conclusion and Discussion\label{conclusion}}

 We investigate the distribution of the final odds in horse races
 organized by the JRA. The distribution of the final odds shows the
 clear power-law $P_o(x)\propto x^{-1}$ at $x<100$. The distribution of
 the winners' odds also shows a power-law behavior, $P_w(x)\propto
 x^{-2}$. In order to explain these distributions, we propose a model
 for the action of bettors.  Numerical simulation of this model shows
 good agreement with the observation.

 These results show that the  actions of the bettors in gambling can be 
described by  our  simple model. In particular, it should be  noted that
  our model does not consider the bettors to be  rational.
 In economics, a rational person  will attempt to maximize his
 expected rewards  and bet on the horse whose expected rewards appear to be
 maximum. In the usual model of the efficient market, the efficiency
  is guaranteed by this rationality. 
 On the other hand, in our model bettors  bet on the horse
 whose winning probability is maximum. Bettors do not bet
 on the weak horse, even if the odds are extremely high. 
 Of course, we may be able to make the model to explain the distribution
 of odds, assuming rational bettors.  For example, the
 different distribution of strength of horses $s_i$ will give the
 different distribution of the final odds. However, such a model will be
 more complicated one  than our model. 
 Therefore it seems natural to conclude that bettors are irrational in
 horse-racing.

 The inconsistency between our investigation and that of Park {\it et
 al.}\cite{Park} is also interesting. They studied the odds
 in Korean horse races and found the power-law distribution of the final
 odds, $P_o(x) \propto x^{-1.7}$. To
 explain this power-law distribution, the assumed that bettors
 overestimate the probability of
 winning of the horse whose odds is small. Due to this effect, bettors
 overestimate the winning probability of strong horse.  Both of their
 empirical data and simulation model are different from ours.
 In Japanese horse races, the distribution of odds shows different
 power-law, $P_o(x)\propto x^{-1}$.
 One of the possible origin of this  difference is  the difference in
 the distribution of the strength of horse. As we noticed, the
 different distribution of the strength of horses will give the different
 distribution of the final odds. However, it will be
 difficult to obtain the distribution of the strength from empirical
 data.  We also note that the model of Parks {\it et al.} cannot explain our
 empirical data even if the distribution of strength is changed. As noticed
 above, their model assumes that the winning probabilities of strong horses are
 overestimated. If it is true, we can obtain larger expected rewards when we
 bet on the horse with large odds. However, our data show that the
 expected rewards is almost independent from odds. 

  Our paper will shed light on the understanding of other
 economic activities of people, such as those in the financial market.
 The dynamics of the financial market is one of the
 main concern in econophysics. In this regards, the understanding of the
 action of  traders plays the crucial role. If we know the action
 of each trader, we can make a reliable model to emulate the
 market. In this paper, we construct a model of bettors in
 gambling market, which can explain the distribution of the final odds
 very well. The knowledge obtained from the study of gambling market
 will give important information to the understanding of the financial market.
\section*{Acknowledgments}

   We are grateful to Takashi Teramoto, Jun-Ichi Fukuda, Yumino Hayase, Makoto Iima and
 Tatsuo Yanagita for the fruitful discussion.


\begin{thebibliography}{99}
 \bibitem{Gabriel}P. E. Gabriel and J. R. Marsden, The Journal of
	 Political  Economy   98 (1990) 874.
 \bibitem{Russo} B. Russo, J. M. Gandar, and R. A. Zuber. Market
	 rationality tests based cross-equation restrictions. Journal of
	 Monetary Economics,  24 (1989) 455.
 \bibitem{Cain} M. Cain, D. Law, and D. Peel, Journal of Forecasting,
	  19 (2000) 575.
 \bibitem{Ioannidis} C. Ioannidis and D. Peel, Economics Letters, 87
	 (2005) 221.
 \bibitem{Park} K. Park and E. Domany, Europhysics Letters, 53 (2001) 419.
\end{thebibliography}



\end{document}